\documentclass[preprint,12pt]{elsarticle}
\usepackage{amsmath,amsfonts}

\usepackage{array}
\usepackage[caption=false,font=normalsize,labelfont=sf,textfont=sf]{subfig}
\usepackage{textcomp}
\usepackage{stfloats}
\usepackage{url}
\usepackage{algorithm}
\usepackage{algpseudocode}
\algnewcommand\algorithmicforeach{\textbf{for each}}
\algdef{S}[FOR]{ForEach}[1]{\algorithmicforeach\ #1\ \algorithmicdo}
\usepackage{booktabs}
\usepackage{verbatim}
\usepackage{graphicx}
\usepackage{balance}
\usepackage[percent]{overpic}
\usepackage{xcolor}
\usepackage{amsthm, empheq}
\usepackage{amsmath}
\usepackage{amssymb}

\usepackage{hyperref}
\usepackage{cleveref}

\newtheorem{theorem}{Theorem}[section]

\newtheorem{lemma}[theorem]{Lemma}
\newtheorem{definition}{Definition}

\usepackage{comment}

\DeclareMathOperator*{\argmin}{argmin}

\usepackage{amssymb}
\usepackage{amsmath}

\journal{arXiv}

\begin{document}

\begin{frontmatter}


\title{Unifying Adaptive Fourier and Möbius-Based Models for Efficient and Interpretable Biomedical Signal Decomposition}

\author[label1]{Christian Canedo\corref{cor1}} 
\author[label2]{Rocío Carratalá-Sáez} 
\author[label1]{Cristina Rueda} 
\cortext[cor1]{Corresponding author: christian.canedo@uva.es - Paseo de Belen, 7 A220 - 983 185876 }

\affiliation[label1]{organization={Department of Statistics and Operations Research, University of Valladolid},
            addressline={Paseo de Belén, 7}, 
            city={Valladolid},
            postcode={47011}, 
            state={Castilla y León},
            country={Spain}}

\affiliation[label2]{organization={Department of Informatics, University of València},
            addressline={Carrer de la Universitat}, 
            city={València},
            postcode={46100}, 
            state={Comunitat Valenciana},
            country={Spain}}
            
\begin{abstract}

Oscillatory biomedical signals such as electrocardiograms (ECG) and electroencephalograms (EEG) call for decompositions that are both computationally efficient and interpretable. This paper establishes a formal connection between two finite-order frameworks that have largely evolved independently: Adaptive Fourier Decomposition (AFD), based on orthonormal Takenaka–Malmquist expansions, and the Frequency-Modulated Möbius (FMM) model, a parametric decomposition built on Möbius transforms with morphologically meaningful parameters. We prove that finite-order AFD and FMM decompositions are mathematically equivalent. Under mild regularity assumptions, we further show that their associated estimation procedures solve the same underlying optimization problem when FMM is formulated with independent Gaussian noise. The results are extended to multi-channel signals, which are central in multilead bioelectric recordings. Practically, the equivalence clarifies how fast AFD approximations (including FFT-based implementations) relate to FMM-style parametrization and component interpretability. We illustrate these implications with an EEG example evaluating approximation behaviour as the number of components increases, and with an ECG use case comparing five-component decompositions on representative beats, contrasting unlabeled AFD components with physiologically identified FMM components. Overall, the proposed equivalence provides a principled basis to leverage the computational advantages of AFD alongside the interpretability of FMM in biomedical signal analysis.



\end{abstract}


\end{frontmatter}

\section{Introduction}
Oscillatory biomedical signals are central to diagnosis, monitoring, and decision support. In electroencephalography (EEG), for instance, rhythmic activity emerges from the superposition of multiple electrical sources and reflects neural dynamics across brain states \cite{cole2017brain}. In electrocardiography (ECG), clinically relevant information is encoded in the morphology and timing of characteristic waveform segments, and subtle alterations can indicate abnormalities in cardiac function. In these and many related settings, the objective is not merely to approximate a signal accurately, but to obtain representations that are robust to noise and morphological variability, computationally efficient for large-scale or repeated analyses, and interpretable at the component level to support downstream biomedical tasks.

Decomposing a signal into simpler constituents is a key step toward interpretation and automated analysis. Spectral decomposition methods are particularly relevant in biomedical contexts \cite{marvsanova2017ecg, roonizi2013morphological}. Classical Fourier Decomposition (FD) remains widely used due to its simplicity, frequency interpretability, and computational efficiency. However, FD can struggle with non-sinusoidal and time-varying oscillations, often producing components that are difficult to relate to physiology. In addition, the global support of Fourier basis functions limits their ability to capture localized temporal features, which is problematic for signals such as the ECG where distinct segments (e.g., P, QRS, T) carry specific physiological meaning.

To address these limitations, more flexible and adaptive decompositions have been developed. Adaptive Fourier Decomposition (AFD) constructs positive-frequency decompositions that accommodate non-sinusoidal components \cite{qianMonoComponents}. It is based on orthonormal expansions using the Takenaka--Malmquist (T--M) system \cite{QianIntrinsic}, enabling compact representations governed by a small number of parameters. From a practical perspective, AFD also benefits from efficient estimation procedures, including FFT-based implementations \cite{QianFFT}, and has been extended toward sparse and complete representations \cite{qian2022sparse, qian2023complete}. These properties make AFD attractive when the primary aim is accurate approximation with favorable computational scaling. AFD has also been applied in diverse domains beyond biomedicine \cite{systems22, games22, forecasting20, series2023, dai2022image, cheng2024deep}. Nevertheless, AFD is fundamentally designed for approximation: its components are not, by construction, intended to correspond to physiologically labeled waveform segments, which can limit interpretability in biomedical workflows.

In parallel, a statistical approach known as Frequency Modulated M\"obius (FMM) decomposition has been proposed \cite{rueda2019frequency}. FMM models a signal as a sum of oscillatory components plus error, where each component corresponds to a scaled M\"obius transformation describing a dominant oscillation. Importantly for biomedical applications, FMM parameters admit direct morphological interpretation, and maximum likelihood estimation is used for parameter inference \cite{rueda2021hidden}. These features have enabled effective use of FMM in the automated interpretation of bioelectric signals such as ECG and EEG, including pattern detection and classification tasks \cite{rueda2022compelling, larriba2023circust, canedo2023novel, rueda2023functional}. In the specific case of ECG analysis, a five-component FMM model can capture the main oscillatory structure of the signal in alignment with the canonical PQRST waves \cite{rueda2022unique}, and recent works have integrated FMM into explainable AI frameworks for heart disease detection using ECG data \cite{yang2022explainable, verardo2023fmm}.

Although AFD and FMM were developed independently and remain relatively separated across communities, they share complex-domain foundations and target the decomposition of non-stationary oscillatory signals. At the same time, they differ in structure and purpose in a way that is highly relevant for biomedical signal processing: AFD provides orthonormal, computationally efficient approximations (including FFT-based acceleration) \cite{QianFFT}, whereas FMM provides waveform-level parametrizations with direct morphological interpretability \cite{rueda2019frequency, rueda2021hidden}. A unified perspective is therefore valuable beyond conceptual clarity. In particular, a formal relationship between both decompositions enables methodological transfer across frameworks: efficient AFD approximation machinery may support FMM-style inference, while FMM parametrization and identification principles motivate more interpretable use of adaptive expansions. This motivates revisiting both methods under a common theoretical and optimization framework.

This paper establishes such a bridge. The main contribution is to prove that finite-order AFD and FMM decompositions are mathematically equivalent and to show that, under mild regularity assumptions, their associated estimation procedures solve the same underlying optimization problem when the FMM model assumes independent Gaussian errors. These results are first developed in the single-channel setting.

However, biomedical recordings are often multi-channel, with shared underlying dynamics observed from different sensors. A representative example is the 12-lead ECG, where each lead records a different projection of the same cardiac activity. Multichannel extensions of both AFD and FMM have been developed to accommodate this setting \cite{QianMultichannel, rueda2022unique}. Building on the single-channel results, we also establish the equivalence of these multichannel extensions. As a practical illustration, we apply both AFD and FMM methods to ECG data, bridging theory with real-world signal analysis and clarifying the complementary roles of fast orthonormal approximation (AFD) and physiologically interpretable parametrization (FMM).

This work does not aim to introduce a new decomposition method. Instead, it shows that AFD and FMM, which have been developed independently in different communities, are mathematically equivalent for finite-order decompositions. This equivalence helps explain empirical similarities reported in the literature and provides a principled foundation to combine the computational advantages of AFD with the interpretability of FMM in biomedical signal analysis.

The remainder of the paper is structured as follows: Section~\ref{Background} introduces the mathematical background. Sections~\ref{Equiv} and~\ref{Equiv2} present the equivalence results for single-channel signals. Section~\ref{3D} extends the framework to the multi-channel setting. Section~\ref{algorithms} describes the estimation algorithms and provides a comparative analysis. Section~\ref{ECG} illustrates the application of both methods to ECG signals, and Section~\ref{Discussion} offers concluding remarks.

\section{Background} \label{Background}

In this section, we provide essential background information necessary for a basic understanding of the paper's methodology, focusing on the representation of signals in the complex domain. 

In Section~\ref{subsec:hardy}, we present the  Analytic Signal (AS) framework to extend real-valued signals to the complex plane and the particular space of analytic functions, the Hardy space $H^2(\mathbb{D}),\mathbb{D}=\{z\in \mathbb{C}: |z| < 1\}$ (\cite{garnett2006bounded, HardySpaces}). In Section~\ref{subsec:collection}, we give definition of interesting elements of  $H^2(\mathbb{D})$, such as the Möbius transforms (\cite{MobiusComplex}), the Blaschke products (\cite{BlaschkeDaniel2, BlaschkeSumerya, BlaschkeSurvey}), and the Takenaka-Malmquist (T-M) system (\cite{mnatsakanyan2022almost, bultheel2003fourier}). These functions are fundamental in establishing a robust signal decomposition framework. 
In Sections~\ref{subsec:AFD} and \ref{subsec:mob} we formulate the AFD and FMM representations, respectively.

\subsection{Analytic signals and Hardy spaces framework} \label{subsec:hardy}

Although researchers often deal with real signals, these signals can be analyzed in the complex domain, which endows the original real functions with interesting properties, such as representation through power series expansions. 

In particular, when the analyzed phenomenon is oscillatory, it is generally accepted that an underlying complex signal exists, even though the imaginary part is not observed. Notably, there are several complex representations that have the observed signal as their real part. The AS is the most widespread of these representations used in the literature.

For a real-valued signal denoted as, 
$\mu(t), t\in[0,2\pi)$, the AS is defined as follows: 

\begin{equation} \label{EQ:AS}
S(t)=\mu(t)+i\mathcal{H}\{\mu(t)\}
\end{equation}
where, $\mathcal{H}$ denotes the Hilbert transform \cite{King_2009}. 

The Hilbert Transform plays a crucial role in this paper, serving as the connection between the real observed signal and the complex space where the representations are formulated. The Hilbert transform gives
the Hilbert space of real-valued functions $L^{2}(\mathbb {R})$ the structure of a complex Hilbert space.

By the change of variable  $z=\exp(it)$, $S(t)$ is also defined as a complex variable function, $S(z), z \in \partial{\mathbb{D}}=\{z\in \mathbb{C}: |z|=1\}$. \\

In the following, we deal with  functions in two spaces, firstly  $H^2(\mathbb{D})$, which denotes the space of all analytic functions with finite norm, as follows:
\begin{equation}
   H^2(\mathbb{D})=\left\{f / \sup_{0<r<1}\left(\int_0^{2\pi}|f(r\exp(it))|^2 dt\right)^{1/2}<\infty\right\}
\end{equation}
and secondly, $L^2(\partial{\mathbb{D}})$, the Hilbert space of square-integrable functions under the inner product:

\begin{equation} \label{EQ:inner}
\langle F(\exp(it)), G(\exp(it)) \rangle=\int_{0}^{2\pi} f(\exp(it))\overline{g(\exp(it))} \frac{dt}{2\pi}.
\end{equation}

Any function of $H^2(\mathbb{D})$ is identifiable with a function in $L^2(\partial{\mathbb{D}})$ through its radial limit $r\to1^{-}$, and vice-versa through the Poisson kernel \cite{rudin1987real}.

\subsection{Basic functions on  \texorpdfstring{$H^2(\mathbb{D})$}{Lg}} \label{subsec:collection}

In this section we  present, on the one hand,  the Blaschke products (the canonical form) and the T-M system, which form the basis for AFD. On the other hand, we present the Möbius transforms, which form the basis for the FMM aproximation.

\begin{definition}  Blaschke product.
    
\textit{Let $\{a_k\}_{k=0}^K$ values in $\mathbb{D}$ and $K\in\mathbb{N}\cup \{\infty\}$.  A  Blaschke product is a function of the form}

$$\mathfrak{B}_{a_0,...,a_{k}}(z) = \prod_{k=0}^K \frac{z-a_k}{1-\overline{a_k}z}.$$\label{EQ:blaschkeprod}
If $K=\infty$, then $\sum_{k=1}^\infty (1-|a_k|)<\infty.$     

\end{definition}

Blaschke products have been extensively studied, with mathematicians establishing their numerous properties and applications across various branches of mathematics \cite{mashreghi2013blaschkeApplications, garcia2018blaschkeConnections}. In our context, these functions are used to construct systems based on monocomponent signals. A monocomponent signal is a complex-valued analytic function with nonnegative frequency components, ensuring physical interpretability \cite{monocomponents}.

\begin{definition} Takenaka-Malmquist system. \label{EQ:TMBasis} 

\textit{The elements of a T-M system $\{B_k\}_{k=0}^K$ are fuctions defined as:}

\begin{equation} 
    B_k(z)=B_{a_0,...,a_{k}}(z) =e_{a_k}(z)\mathfrak{B}_{a_0,...,a_{k-1}}(z), 
\end{equation}

where for $ a_k\in \mathbb{D}, k=0,\dots,K$, for $K\in\mathbb{N}\cup\{\infty\}$,

$\mathfrak{B}_{a_0,...,a_{k-1}}(z)$ is a Blaschke product and

\begin{equation} \label{EQ:szego}
    e_{a_k}(z) = \frac{\sqrt{1-|a_k|^2}}{1-\overline{a_k}z}, 
\end{equation}

is known as the normalised Szëgo kernel.

\end{definition}
$\mathfrak{B}_{a_0,...,a_{k}}(z)$ is a bounded analytic function in $\mathbb{D}$ constructed to have zeroes at $a_0,...,a_k$ and  $B_{a_0,...,a_{k}}(z)$ is the normalized version. Moreover, for any $k>0$, $B_{a_0,...,a_{k-1}}(z)$ and $B_{a_0,...,a_k}(z),$ are orthogonal under the inner product defined in \eqref{EQ:inner} and therefore T-M is an  orthonormal system.
\\

The T-M systems are  bases of the $H^p(\mathbb{D})$ spaces $1<p<\infty$ in the sense that a function $f(z) \in H^p(\mathbb{D})$ can be represented through a T-M basis as

\begin{equation}
    \sum_{k=0}^K\langle f(z),B_{a_0,...,a_k}(z)\rangle B_{a_0,...,a_k}(z) \overset{K\rightarrow{\infty}}{\longrightarrow} f(z).
\end{equation}
This result has been proved in \cite{coifman2019phase} and was previously obtained for $H^2(\mathbb{D})$ \cite{qian2014convergence}, which is the particular case in which we are interested. 

The T-M system offers greater flexibility compared to other systems, such as the classical Fourier basis or the Laguerre system. This flexibility arises because the zeros of the T-M system are distributed throughout the entire unit disk. In contrast, the Fourier basis has all its zeros concentrated at the origin of the complex plane, and the zeros of the Laguerre system are located on the real line. This distribution of zeros leads to a more efficient decomposition of functions than the other systems.\\

 When $a_0=0$, $B_{a_0,...a_k}(z), K>0$ is a mono-component  \cite{monocomponents}. Besides, the individual factors of the Blaschke products are particular cases of the Möbius transform which is defined below. 

\begin{definition} Möbius transform. 

\textit{Let $a\in\mathbb{D}$ and $\varphi\in\mathbb{R}$, the Möbius transform is defined as},
\begin{equation} \label{EQ:mobius}
    m_{a}(z)=\frac{z-a}{1-\bar{a}z}, z\in\mathbb{D}.
\end{equation}

\end{definition}

This function, can be also be defined in $\partial{\mathbb{D}}$, in fact, is the simplest transformation that maps $\partial{\mathbb{D}} \to \partial{\mathbb{D}}$. Moreover, it has been proved that $ m_{a,\varphi}(z), a \in \partial{\mathbb{D}}$ is a mono-component that describes up-down-up or, equivalently, single oscillation signals~\cite{rueda2019frequency}.  
\\

\subsection{Adaptive Fourier Decomposition} \label{subsec:AFD}

The finite AFD of a function $f(z)\in H^2(\mathbb{D})$ is  based on a T-M system $\{B_k(z)\}_{k=0}^K$ as follows,

\begin{equation} \label{EQ:AFDModel}
    f(z)=c_0+\sum_{k=1}^K c_kB_k(z), \,
\end{equation}

where $z\in\mathbb{D}$, $a_0=0$, $a_k,\, k>0$ are the zeros of $f(z)$ and $c_k=\langle f(z),B_k(z) \rangle\in \mathbb{C}, k>0$.

\subsection{Frequency Modulated Möbius} \label{subsec:mob}

A single FMM complex signal is defined as a real-variable function as: 

\begin{equation} \label{EQ:FMMModel}
    S(t)=\phi_0+\phi_1m_{a}(\exp(it)), t \in [0,2\pi]
\end{equation}

where $m_{a}(z)$ is a Möbius transform in $\partial{\mathbb{D}}$ and $\phi_0, \phi_1\in\mathbb{C}$. 

The real part  of an  FMM  complex signal with $\phi_0=0$ and $|\phi_1|=1$ is called an FMM wave:

\begin{definition} FMM Wave

\begin{equation} \label{EQ:MobiusWave}
    W(t; \alpha, \beta, \omega) = \cos(\beta+\Phi(t;\alpha, \omega))
\end{equation}
\end{definition}

where $\Phi(t; \alpha,\omega)= 2\arctan\left(\omega\left(\tan\left(\frac{t-\alpha}{2}\right)\right)\right)$, $\alpha,\beta\in[0,2\pi)$ and $\omega\in(0,1]$.
This wave is bounded between $[-1,1]$ and it depends on parameters $\alpha$, $\beta$ and $\omega$. These parameters are directly interpreted: $\alpha$ locates the oscillation along $t\in[0,2\pi)$, $\omega$ measures the width of the oscillation, and the peak-through pattern is described by $\beta$.

$\alpha$ and $\omega$ are related to the polar coordinates of the parameter $a\in\mathbb{D}$ defined in the standard Möbius formulation given in \eqref{EQ:mobius}, as follows:

\begin{equation} \label{EQ:omegaRelation}
   \arg(a)=(\alpha+\pi),\     |a|=\frac{1-\omega}{1+\omega} 
\end{equation}

besides, $\beta=\varphi+\alpha$.

An FMM$_K$ complex signal with K components is defined as: 

\begin{equation} \label{EQ:FMM_KModel}
    S(t)=\phi_0+\sum_{k=1}^K\phi_km_{a_k}(\exp(it)), t \in [0,2\pi)
\end{equation}
and, the corresponding FMM$_K$ real signal is defined as, 
\begin{equation}\label {EQ:ReFMMk}
Re(S(t))=M+\sum_{k=1}^K A_k W(t;\alpha_k,\beta_k,\omega_k), t \in [0,2\pi)
\end{equation}

Now, from well-known trigonometric relationships, $Re(S(t))$ can be written as follows,

\begin{equation} \label{EQ:FMMlinear}
    Re(S(t))=M+\sum_ {k=1}^K A_k\cos(\beta_k)\cos(\Phi(t;\alpha_k,\omega_k))-\sum_{k=1}^K A_k\sin(\beta_k)\sin(\Phi(t;\alpha_k,\omega_k))
\end{equation}

In this additive nonlinear representation, the parameters have different roles, and the estimation problem is addressed differently, as we will see in the next sections. On the one hand, $(\alpha_k,\omega_k), k=1,...,K$, are the nonlinear parameters, while $M, \delta_k= A_k\cos(\beta_k)$ and $ \gamma_k=-A_k\sin(\beta_k)$ are the linear ones. The new parametrization in terms of $\delta$ and $\gamma$ will be used below to simplify the expressions.


\section{Finite-order equivalence and component-level interpretation}\label{Equiv}
\subsection{Equivalence theorem and parameter correspondence}\label{Equiv:theorem}

In this section, we prove the main result of this paper, Theorem~\ref{Theorema},  which demonstrates the equivalence between the AFD and FMM decompositions for a given complex signal $F(z)$ in $\partial {\mathbb{D}}$ defined with an AFD with a finite number of terms, $ K<\infty$. Furthermore, Theorem~\ref{Theorema} provides a matrix expression that relates FMM parameters to AFD parameters.

The key step to proving Theorem~\ref{Theorema} is to express the Blaschke product as a linear combination of Möbius functions, accomplished with the aid of two lemmas. Lemma~\ref{lma1} addresses the order-2 case, while Lemma~\ref{lman} covers the general case. The proofs of the lemmas and the theorem are provided in the supplementary material.

\begin{lemma} \label{lma1} 

\textit{Let $m_{a_1}(z)$ and $m_{a_2}(z)$ be two Möbius transforms with $a_1\neq a_2$. Then, the product $m_{a_1}(z)m_{a_2}(z)$ can be expressed as a linear combination $m_{a_1}(z)m_{a_2}(z) = \lambda_0+\lambda_1m_{a_1}(z)+\lambda_2m_{a_2}(z)$, where:}

\begin{equation} \label{EQ:betas}
\lambda_0 = \frac{a_1-a_2}{\overline{a_1}-\overline{a_2}},\,\,\lambda_1 = \frac{1-\overline{a_1} a_2}{\overline{a_1}  - \overline{a_2}},\,\,\lambda_2 = \frac{1-\overline{a_2}a_1}{\overline{a_2} - \overline{a_1}}.
\end{equation}

\end{lemma}

\begin{lemma} \label{lman} 

\textit{Let $m_{a_k}(z)$ be a Möbius transform $(z-a_k)/(1-\overline{a_k}z)$, $k=1,...,K$ where $a_k\neq a_{k^\prime}$, $k\neq k^\prime$. Then, a Blaschke product can be expressed as follows}

\begin{equation} \label{EQ:ProductnMobius}
  \mathfrak{B}_{a_1,...,a_K}(z) = \xi_0^K+\sum_{k=1}^K\xi_k^Km_{a_k}(z),
\end{equation}

\textit{where}

\begin{equation} \label{EQ:ProductnMobius1}
\xi_0^k = \sum_{i=1}^{k-1}\xi_i^{k-1}\lambda_0^{(i,k)}
\end{equation}
\begin{equation} \label{EQ:ProductnMobius2}
\xi_i^k = \xi_i^{k-1}\lambda_1^{(i,k)}, i = 1,...,k-1
\end{equation}
\begin{equation} \label{EQ:ProductnMobius3}
\xi_k^k = \xi_0^{k-1} + \sum_{i=1}^{k-1}\xi_i^{k-1}\lambda_2^{(i,k)}
\end{equation}

\textit{for k = 1,...,K and $\xi_0^1 = 0, \xi_1^1 = 1$, and where $\lambda_0^{(i,k)}, \lambda_1^{(i,k)}$ and $\lambda_2^{(i,k)}$ are the coefficients obtained by \eqref{EQ:betas} given $a_i$ and $a_k$.}

\end{lemma}

\begin{theorem} \label{Theorema} \textit{Let $a_1 \neq \cdots \neq a_K \in \mathbb{D}$, and let $\{B_{k}\}_{k=0}^K$ and $\{m_{a_k}\}_{k=0}^K$ be the associated Takenaka-Malmquist and Möbius bases, respectively. Then, for every $\boldsymbol{c}=(c_0,c_1,...,c_K)'\in \mathbb{C}^{K+1}$, there exists a unique $\boldsymbol{\phi}=(\phi_0,\phi_1,...,\phi_K)' \in \mathbb{C}^{K+1}$ such that}
$$
F(z) = c_0 + \sum_{k=1}^K c_k B_k(t) = \phi_0 + \sum_{k=1}^K \phi_k m_{a_k}(t) \in L_{}^2(\partial\mathbb{D}),
$$
\textit{ where $\boldsymbol{\phi} = \mathcal{M}_{\boldsymbol{a}} \, \boldsymbol{c}$, and}

\begin{equation*} \mathcal{M}_{\boldsymbol{a}}= \begin{pmatrix}
1 & \frac{a_1}{\sqrt{1-|a_1|^2}} & \dots & \frac{a_k\xi_0^{k-1}+\xi_0^k}{\sqrt{1-|a_k|^2}} & \dots &  \frac{a_K\xi_0^{k-1}+\xi_0^k}{\sqrt{1-|a_k|^2}}\\
0 & \frac{1}{\sqrt{1-|a_1|^2}} & \dots & \frac{\xi_1^k\sqrt{1-|a_k|^2}}{\overline{a_1}-\overline{a_k}} & \dots &  \frac{\xi_1^K\sqrt{1-|a_K|^2}}{\overline{a_1}-\overline{a_K}}\\
\vdots & \vdots & \ddots & \vdots & \ddots & \vdots\\
0 & 0 & \dots & \frac{\xi_k^k}{\sqrt{1-|a_k|^2}} & \dots & \frac{\xi_k^K\sqrt{1-|a_K|^2}}{\overline{a_k}-\overline{a_K}}\\
\vdots & \vdots & \ddots & \vdots & \ddots & \vdots\\
0 & 0 & \dots & 0 & \dots & \frac{\xi_K^K}{\sqrt{1-|a_K|^2}}
\end{pmatrix}  ,
\end{equation*}

\textit{where the values of $\xi_k^{j}, k=1,...,K, j<k$ are given by equations \eqref{EQ:ProductnMobius1}, \eqref{EQ:ProductnMobius2} and \eqref{EQ:ProductnMobius3}. Furthermore, since the change of basis matrix is invertible, it follows reciprocally that for every $\boldsymbol{\phi} \in \mathbb{C}^K$, there exists a unique 
 $\boldsymbol{c}\in\mathbb{C}^K$, given by $\boldsymbol{c}=\mathcal{M}^{-1}_{\boldsymbol{a}}\boldsymbol{\phi}$.} 
    
\end{theorem}

Theorem~\ref{Theorema} shows, firstly, that the parameters $(a_1,\dots,a_K)$ are common to AFD and FMM, and using \eqref{EQ:omegaRelation} can be reparametrized as $(\alpha_1,\omega_1,\dots,\alpha_K,\omega_K)$. Secondly, it demonstrates that the AFD coefficients and the linear FMM parameters can be transformed into each other through the change-of-basis matrix $\mathcal{M}$. \\

\subsection{Component-level differences and interpretation} \label{Equiv:interpretation}

Despite the equivalence proved in Theorem~3.3, the two representations differ in how the parameters $a_1,\ldots,a_K$ contribute to individual components. On the one hand, the $k$-th AFD term $c_k B_k$ depends on the whole sequence $(a_1,\ldots,a_k)$, whereas the $k$-th FMM term depends only on $a_k$. As a consequence, an AFD component may reflect the accumulation of several previously selected oscillatory structures, while an FMM component is associated with a single dominant oscillation.

On the other hand, the order of $(a_1,\ldots,a_K)$ affects AFD components in two ways: through the induced T--M basis functions and through the coefficients $c_k$. In contrast, the FMM parameters are invariant to permutations of the $a$'s. This order-dependence can complicate the practical interpretation of AFD, particularly in noisy settings, since it introduces additional variability in the shapes of individual components.

Next, to illustrate the relationship between both representations and the impact of ordering, we consider the case $K=2$, as $K=1$ reduces to a change of scale. The equivalence for $K=2$ is established as follows:

\begin{equation} \label{EQ:phi0}
\phi_0 = c_0 + \frac{c_1}{\sqrt{1-|a_1|^2}}a_1 + \frac{c_2}{\sqrt{1-|a_2|^2}}\frac{a_1-a_2}{\overline{a_1}-\overline{a_2}},
\end{equation}

\begin{equation} \label{EQ:phi1}
\phi_1 = \frac{c_1}{\sqrt{1-|a_1|^2}} + c_2\frac{\sqrt{1-|a_2|^2}}{\overline{a_1}-\overline{a_2}},
\end{equation}

\begin{equation} \label{EQ:phi2}
\phi_2 = \frac{c_2}{\sqrt{1-|a_2|^2}} \frac{1-\overline{a_2}a_1}{\overline{a_2}-\overline{a_1}},
\end{equation}

where $c_0, \phi_0\in\mathbb{R}$, $c_1, c_2,\phi_1,\phi_2\in \mathbb{C}$ and $a_1\neq a_2$. \\

\begin{figure*}[ht!] 
    \centering
    \includegraphics[width=\textwidth]{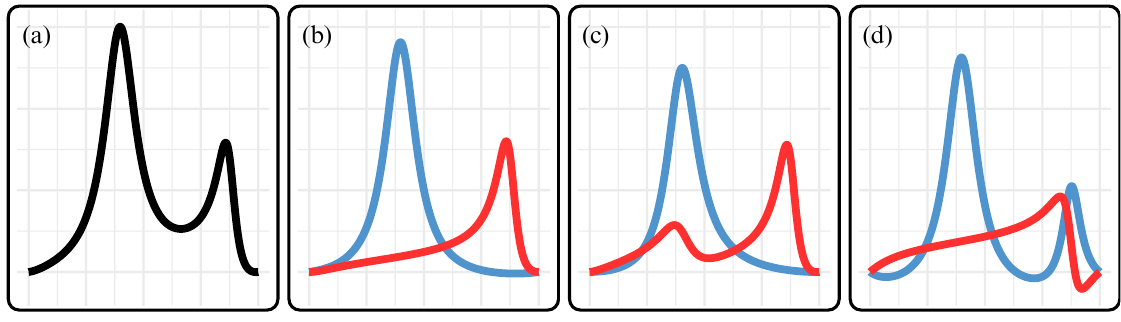}
    \caption{Illustration of FMM decomposition and AFD for   $S(t)=c_1+c_1B_1(t)+c_2B_2(t)=\phi_0+\phi_1m_{a_1}(t)+\phi_2m_{a_2}(t)$ where $a_1=0.6e^{2.5i}, a_2=0.74e^{5.5i}$ and $c_0=0, c_1=2e^{3.65i}, c_2=e^{2.15i}$ and $\phi_0,\phi_1$ and $\phi_2$ are obtained from \eqref{EQ:phi0}, \eqref{EQ:phi1} and \eqref{EQ:phi2}, respectively. (a) Represents the real part of $S(t)$, (b) represents the real part of $\phi_1m_{a_1}(t)$ (blue) and $\phi_2m_{a_2}(t)$ (red), (c) represents the real part of $c_1B_1(t;a_1)$ (blue) and $c_2B_2(t;a_1,a_2)$ (red), (d) represents the real part of $c_1B_1(t;a_2)$ (blue) and $c_2B_2(t;a_2,a_1)$ (red).} \label{fig:changebasis}
\end{figure*}

As a toy example, consider an AFD signal with parameters $a_1 = 0.6 e^{2.5 i}$, $a_2 = 0.74 e^{5.5 i}$ and $c_0 = 0$, $c_1 = 2 e^{3.65 i}$, $c_2 = e^{2.15 i}$. Figure~1(a) shows the real part of the signal, where two prominent peaks suggest two dominant components. Figure~1(b)--(c) display the real parts of the corresponding FMM and AFD components, respectively, and Figure~1(d) shows the AFD components after interchanging the roles of $a_1$ and $a_2$. The figure highlights that each FMM component delineates a single peak, whereas AFD components may mix the visible structures and change when the order of the $a$'s is modified.

\section{Equivalent optimization problems for AFD and FMM: a statistical perspective } \label{Equiv2}

The development of AFD and FMM has occurred independently, and the estimation problem has also been approached differently in each case. FMM has been presented as a statistical model with the main objective of estimating key features of real-valued signals. It assumes an additive signal plus Gaussian error model, and the likelihood approach has been adopted to solve the estimation problem. The independence of errors is often assumed, although the case of correlated errors has also been considered (\cite{canedo2023novel}). It has been particularly effective in analyzing bioelectric signals, such as EEG and ECG, identifying patterns, and classifying anomalies (\cite{rueda2022compelling}).  The number of components, $K$, is fixed in each application and corresponds to the well-known, physiologically understandable, prominent oscillations of the signal. 

In contrast, AFD has focused more, at least initially, on theoretical properties. It has been presented as a greedy method for decomposing signals by adaptively choosing the parameters of a T-M system \cite{QianCore}. 

When used to approximate a real function, the goal is to minimize the variability of the residuals as much as possible, given a threshold, often resulting in a high and variable number of components. In applications, it has primarily been used for signal compression and denoising \cite{ECGdenoisingWang, compressionECGma2014novel, wang2014muscle} .\\ 

In this section, we deal with the decomposition of an observed real-valued signal into a fixed number of components. We demonstrate the equivalence between the AFD and FMM estimation problems, the latter assuming independent Gaussian errors. \newline

Let $\boldsymbol{t}=(t_1,...,t_n)'$ where $0\le t_1 \le....\le t_n \le 2\pi$ and
 $X(\boldsymbol{t})=(X(t_1),...,X(t_n))'$ be the observed  real-valued data. 

On the one hand, the FMM approach assumes that $X(\boldsymbol{t})$ is an FMM\textsubscript{K} signal plus a random noise as follows:

$$X(\boldsymbol{t}) = \mu_K(\boldsymbol{t},\theta) + \epsilon(\boldsymbol{t})$$
$$\mu_K(\boldsymbol{t},\theta) = M+\sum_{k=1}^K A_k W(\boldsymbol{t};\alpha_k,\beta_k,\omega_k)= $$
$$ = M+\sum_ {k=1}^K [\delta_k\cos(\Phi(\boldsymbol{t};\alpha_k,\omega_k)) + \gamma_k\sin(\Phi(\boldsymbol{t};\alpha_k,\omega_k))]$$

where $\epsilon(\boldsymbol{t})\sim N(0,\sigma)$, $\theta$ denotes the complete vector of parameters with parameter space $\Theta$. A maximum-likelihood approach is adopted to estimate $\theta\in\Theta$. Following a statistical notation, we use a hat over a parameter (or vector of parameters) to denote its estimation. Then, the FMM estimation problem is defined as follows,

\begin{equation} \label{EQ:minFMM}
    \hat{\theta}=\argmin_{\theta\in\Theta} \|r(\boldsymbol{t},\theta)\|^2
\end{equation}

where $r(\boldsymbol{t},\theta)=X(\boldsymbol{t})-\mu_K(\boldsymbol{t},\theta)$ are real-valued residuals. \\

On the other hand, the AFD approach starts from the AS associated to $X(\boldsymbol{t})$ denoted by $Z(\boldsymbol{t})$, instead of from $X(\boldsymbol{t})$. Then, the AFD estimation problem is formulated as follows: 

\begin{equation}  \label{EQ:minAFD}
    \hat{\upsilon} = \argmin_{\upsilon\in\Upsilon} \|r^*(\boldsymbol{t},\upsilon)\|^2 
\end{equation}

where $r^*(\boldsymbol{t},\upsilon)=Z(\boldsymbol{t})-S_K(\boldsymbol{t},\upsilon)$ are residuals in the complex space, $\upsilon$ denotes the complete vector of parameters and $\Upsilon$ is the parameter space. \\

In the following, we prove the equivalence between \eqref{EQ:minAFD} and \eqref{EQ:minFMM}. For simplicity of notation, and without loss of generality, we assume that the signal has zero mean ($c_0=0$).

First, it is well established in the literature \cite{monocomponents} that : 
\begin{equation} \label{EQ:Monocomp}
  Im(c_kB_k^a(t))=\mathcal{H}(Re(c_kB_k^a(t))) 
\end{equation}

Then,

$$\|r^*(\boldsymbol{t},\upsilon)\|^2=$$
$$=\|X(\boldsymbol{t})+ i\mathcal{H}\{X(\boldsymbol{t})\}-Re(S_K(\boldsymbol{t},\upsilon)) -i\mathcal{H}\{Re(S_K(\boldsymbol{t},\upsilon)\} \|^2=$$
$$=\|X(\boldsymbol{t})-Re(S_Kk(\boldsymbol{t},\upsilon)) + i\mathcal{H}\{X(\boldsymbol{t})-Re(S_K(\boldsymbol{t},\upsilon))\}\|^2=$$ 
$$=2\|X(\boldsymbol{t})-Re(S_K(\boldsymbol{t},\upsilon))\|^2.$$

Where, the first and second equalities follows from \eqref{EQ:Monocomp} and the linearity of the Hilbert transform, and the third equality follows from the norm-preserving property of the Hilbert transform.

Now, let $\boldsymbol{\phi}=\mathcal{M}_{\boldsymbol{a}}\,\boldsymbol{c}$, then from  the Theorem~\ref{Theorema} follows that

$$ Re \left( c_0+ \sum_{k=1}^Kc_kB_k^a(\boldsymbol{t})\right) = Re\left(\phi_0+\sum_{k=1}^K\phi_km_{a_k}(\boldsymbol{t})\right),$$
Now, from \eqref{EQ:ReFMMk} it follows that,

$$Re\left(\phi_0+ \sum_{k=1}^K\phi_km_{a_k}(\boldsymbol{t})\right) =M+\sum_{k=1}^K A_k W(\boldsymbol{t};\alpha_k,\beta_k,\omega_k)$$

and the equivalence between the optimization problem follows.

\section{Multichannel decompositions }\label{3D}

In this section, we discuss the multi-channel case as a natural extension of the previous single-channel study. Addressing multi-channel signals is crucial, as they often provide more comprehensive information by capturing an electrophysiological event from various channels. A clear example is the ECG, where the heart's electrical activity is recorded using multiple electrodes.

Simply applying methods separately to each channel neglects the shared information between channels, resulting in physically meaningless estimates and necessitating more complex models, which can reduce computational efficiency. In fact, univariate approaches may produce varying numbers of misaligned components across different channels. 
This issue is significantly mitigated by assuming the equality of   $a_k$ or $(a_k,\omega_k)$  across channels. This responds to the idea that the channels are different observers of the same events from different perspectives. Thus, the components might share some features (parameters) across channels.

Specifically, the multi-channel FMM, the 3DFMM model,  defined with $(a_k,\omega_k)$  equal across channels, represents the individual signals as projections of a vector that follows a circular trajectory in a plane \cite{rueda2022unique}. In this way, the nonlinear parameters describe this trajectory, while the linear parameters describe the direction and the intensity of observation. This interpretation is particularly suitable for bioelectric signals. The nonlinear parameters are crucial for describing the phenomena at hand, as demonstrated in previous works \cite{rueda2019frequency, rueda2021IEEE, rueda2022unique}.

Furthermore, the multi-channel AFD decomposition assumes the same T-M base across channels, although the justification is less explored \cite{QianMultichannel}.
\\

The equality of $a_k$ or $(a_k,\omega_k)$ across channels is the main characteristic and common advantage of the FMM and AFD approaches for analyzing multi-channel oscillatory signals.
From this, the equivalence of the decompositions in the multi-channel case is derived straightforwardly. Similarly, the equivalence between the multi-channel FMM and AFD optimization problems directly follows from the single-channel case, as the objective functions are defined as the sum of the corresponding functions for individual channels.
\section{Estimation algorithms and numerical experiments} \label{algorithms}

In this section, we discuss the algorithms proposed by various researchers for the independently developed FMM and AFD methodologies. While the algorithms are highly similar, reflecting the equivalence between the two methods as demonstrated in earlier sections, each possesses distinctive characteristics. The key similarities and differences between the algorithms are examined here. Finally, we present two numerical experiments to illustrate the performance of the algorithms.\newline

The algorithms are specifically designed to focus exclusively on optimizing the non-linear parameters $a_k$ or, equivalently, $(\alpha_k,\omega_k)$. This is achieved by expressing the linear parameters in terms of the non-linear ones, as explained below. \newline

Firstly, the linear FMM parameters are obtained by solving the normal equations:
\begin{equation}\label{opt1} (\hat M,\hat \delta_1,\hat \gamma_1,\dots,\hat \delta_K,\hat \gamma_K)'=(\boldsymbol{M}'\boldsymbol{M})^{-1}\boldsymbol{M}'X(\boldsymbol{t}), 
\end{equation}
where $\boldsymbol{M}$ is a design matrix that depends on $\alpha_k,\omega_k$. $\boldsymbol{M}$ has $2K+1$ columns, consisting of a column of ones ($n\times1$) for the intercept term $M$, and $K$ blocks, each containing two columns: 
$$[\cos(\Phi(\boldsymbol{t};\alpha_k,\omega_k)),\, \,\sin(\Phi(\boldsymbol{t};\alpha_k,\omega_k)],$$ corresponding to the coefficients $(\delta_k,\gamma_k)$.\newline

Secondly, for fixed values $a_1,...,a_K$, the AFD coefficients are computed as:

\begin{equation}\label{opt2}
\hat c_k = \langle H_{k}(\boldsymbol{t}),e_{a_k}(\boldsymbol{t})\rangle=\sum_{j=1}^n H_{k}(t_j)\overline{e_{a_k}(t_j)},
\end{equation}

where $e_{a_k}$ is the Szëgo kernel defined in \eqref{EQ:szego} and $H_{k}(\boldsymbol{t})$ is the $k$-th reduced reminder, given by:

\begin{equation}\label{opt3}
H_{k}(\boldsymbol{t})=(H_{k-1}(\boldsymbol{t})-\hat c_{k-1}e_{a_k}(\boldsymbol{t}))\frac{1-\Bar{a}_ke^{i\boldsymbol{t}}}{e^{i\boldsymbol{t}}-a_k},
\end{equation}

with the initial remainder defined as $H_{1}(\boldsymbol{t})=(Z(\boldsymbol{t})-\hat c_0)/e^{i\boldsymbol{t}}$, where $\hat c_0$ is the empirical mean of the original signal.  \newline

The complexity of the estimation problem grows exponentially with \( K \), making it computationally infeasible to find the optimal solution. As a result, all algorithms adopt an iterative approach, estimating each \( a_k \) individually in sequence. 

To manage this process, the algorithms evaluate a grid of candidate values at each step to identify the optimal parameters. From Theorem~\ref{Theorema} and equation \eqref{EQ:omegaRelation}, we have shown that \( \alpha_k \) and \( \omega_k \) correspond to the polar coordinates of \( a_k \). Notably, the parameter space defined by \( (\alpha_k, \omega_k) \) is identical to the space of polar coordinates for \( a_k \). Therefore, the grid constructed for \( (\alpha_k, \omega_k) \) also serves as a grid for \( a_k \), which can be defined as:

\[
\mathcal{D} = \{ \alpha_i \in [0, 2\pi), i = 1, \dots, N_\alpha; \, \omega_i \in [0, 1), i = 1, \dots, N_\omega \},
\]

where \( N_\alpha, N_\omega \in \mathbb{N} \) denote the number of values considered for \( \alpha_k \) and \( \omega_k \), respectively.

The core algorithmic scheme for both methodologies involves iteratively selecting the optimal \( a_k \in \mathcal{D} \), given the residuals produced by the previous estimations of \( a_1, \dots, a_{k-1} \), which are treated as fixed. In step \( k \), we determine \( a_k \in \mathcal{D} \) by solving the optimization problem:

\begin{equation} \label{eq:parOptAFD}
\argmin_{a_k \in \mathcal{D}} \left| Z(\boldsymbol{t}) - \left( c_0 + \sum_{j=1}^{k-1} \widehat{c_j} \widehat{B}_j(\boldsymbol{t}) + c_k B_k(\boldsymbol{t}) \right) \right|^2,
\end{equation}

where each \( \widehat{c_j} \) is determined by \( \widehat{B_j} \), as expressed in equation \eqref{opt2}. The orthonormality of the AFD components simplifies the computation significantly, reducing the previous problem to the equivalent form:

\begin{equation}
\argmin_{a_k \in \mathcal{D}} \left| \langle H_k(\boldsymbol{t}), e_{a_k}(\boldsymbol{t}) \rangle \right|^2,
\end{equation}

where \( H_k(\boldsymbol{t}) \) is the \( k \)-th reduced remainder, defined in equation \eqref{opt3}. A notable improvement is the incorporation of the FFT algorithms to approximate $|\langle H_{k-1}, e_{a_k}\rangle|$ as it is shown in \cite{QianFFT}. This approach substantially reduces computational time. \newline

In contrast, the FMM model involves more intricate computations. As the FMM components lack orthogonality, solving the equivalent problem in equation \eqref{eq:parOptAFD} requires recalculating all previous linear parameters \( (\delta_1, \gamma_1, \dots, \delta_{k-1}, \gamma_{k-1}) \) at each step \( k \) through the normal equations in \eqref{opt1}, making the process significantly more computationally demanding. To address this, a simplification is adopted in which only \( (\delta_k, \gamma_k) \) are estimated. This distinction represents the main difference between the FMM and AFD approaches as proposed in the literature.

Moreover, the FMM algorithm optimizes parameters outside the grid by using a continuous maximum likelihood approach. The original FMM algorithm further iterates through a backfitting procedure\cite{FMMpackage}, detailed in the supplementary material. The two AFD algorithms, Core AFD and Cyclic AFD \cite{QianCore, QianCyclic, qian2019enhancement}, which follow a similar approach, are also described in the supplementary material. The performance of these algorithms is compared numerically in the following section.

\subsection{Experimental Comparison of Optimization Methods}  \label{subsec:F}

We present two experiments to evaluate the performance of the optimization algorithms. Two key aspects are evaluated respectively in both experiments: the fast energy convergence property, verified in comparison to the Fourier basis in Experiment 1; and  the convergence to a local minimum when sufficient iterations are performed by the algorithms, in Experiment 2. \newline 

\begin{figure*}
\centering 
\includegraphics[width=\textwidth]{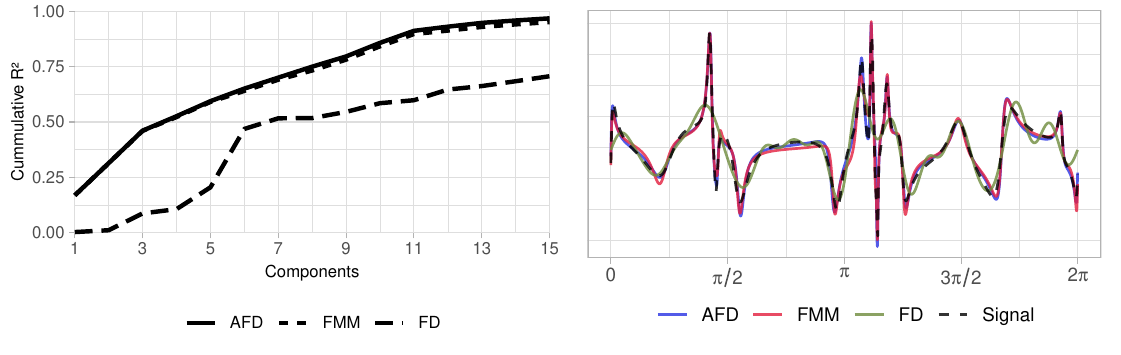} 
\caption{\textbf{Results of the Experiment 1.} The original EEG signal corresponds to Patient 316 from the ICARE Database \cite{Amorim2023ICARE} (Hour 26, Channel Fp1-F7). Panel (a) shows the cumulative $R^2$ measure when each component $k$ is fitted over the residuals of the previous $k-1$ components, which remain fixed. Panel (b) shows the fitted signal with $15$ components to the given signal.} 
\label{fig:experiment1} 
\end{figure*}

\textbf{Experiment 1}:

An FMM model with 15 components has been fitted to a real 10-second EEG signal from a comatose patient in the ICARE database \cite{Amorim2023ICARE}. A total of 1000 equally spaced observations of the signal have been sampled. Our focus is on the predictive performance of AFD/FMM, particularly in the iterative component search. To ensure a fair comparison, we evaluate the AFD and FMM algorithms using a single backfitting iteration and no post-optimization. The search space is restricted to a grid $\mathcal{D}$ with dimensions $N_\alpha = 1000$ and $N_\omega = 50$.

Additionally, both methods are compared to an FD model. Figure \ref{fig:experiment1}, panel (a), shows the cumulative $R^2$ progression for the core AFD, FMM, and FD algorithms. The $R^2_K$ measure is defined as follows:

\begin{equation}
    R^2_K = 1 - \frac{\sum_{j=1}^n (X(t_j) - \hat\mu_K(t_j))^2}{\sum_{j=1}^n (X(t_j) - \Bar{X})^2},
\end{equation}

where $\hat\mu_K$ denotes the predicted signal of any model with $K$ components.

The core AFD algorithm extracts variability slightly faster than the FMM algorithm, with both being significantly faster than the FD. Figure \ref{fig:experiment1}, panel (b), presents the predicted signals using AFD, FMM, and FD with 15 components. AFD and FMM demonstrate excellent adaptability to the signal, particularly in capturing sharper peaks.
\begin{figure*}
\centering 
\includegraphics[width=\textwidth]{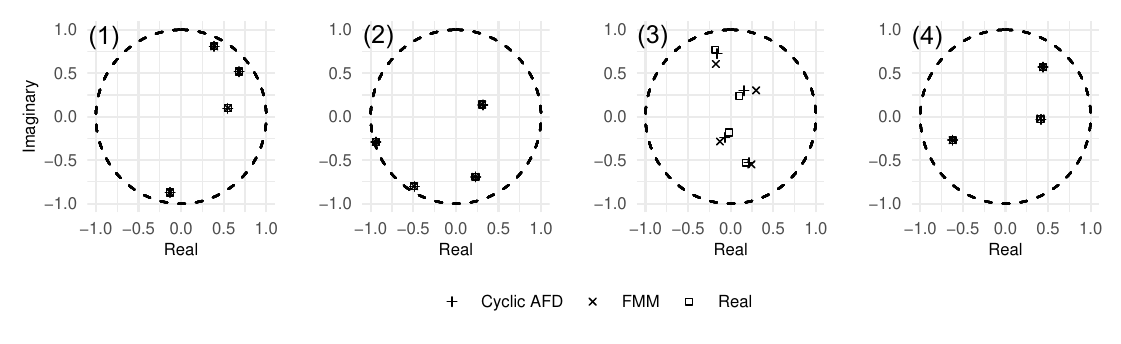} 
\caption{\textbf{Results of the Experiment 2.} Examples 1-4 from \cite{QianCyclic}: Each subplot represents the true and the estimated $a$ values using Cyclic AFD and FMM backfitting algorithms. The dashed line represents the unit circle ($\partial\mathbb{D}$), which is the bound for the $a$ parameters.} 
\label{fig:experiment2} 
\end{figure*}

\textbf{Experiment 2}: To investigate the convergence behavior of the FMM backfitting and Cyclic AFD algorithms, we consider the same synthetic signals from the original Cyclic AFD paper \cite{QianCyclic}, referred to as Examples 1 to 4. For the FMM optimization algorithm, we use a grid with dimensions $N_\alpha = 100$ and $N_\omega = 36$, along with a post-optimization step using the Nelder-Mead method. For the AFD optimization algorithm, a denser grid with $N_\alpha = 1000$ and $N_\omega = 1000$ is employed to prevent numerical resolution issues.

True and estimated  $|a_k|$ values for the four examples are represented in Figure \ref{fig:experiment2}.  Examples 1, 2, and 4 correspond to signals with $|a_k|$ values moderately distant from zero, meaning they deviate significantly from a sinusoidal pattern. As illustrated in the figure, both algorithms successfully converge to the true parameter values in these cases. In contrast, Example 3 presents a more challenging scenario, featuring two components that are nearly sinusoidal ($|a_k|$ close to 0). In this case, the algorithms struggle to estimate the parameters with the same level of accuracy as in the previous examples, with AFD providing slightly more accurate parameter estimates than FMM. Both AFD and FMM were developed to handle signals with pronounced peaks, so it is not surprising that they perform poorly in this scenario.

\section{Application to ECG: five-component decomposition and wave identification}\label{ECG}
The ECG is a graphical record of the heart’s electrical activity, which originates spontaneously in the heart and propagates periodically.

The timing and morphology of the ECG provide valuable information about cardiac function, making it one of the most widely used noninvasive techniques for diagnosing heart disease \cite{rafie2021ecg, ikeda2021ecg}. Automatic ECG interpretation can support clinical decision-making for diagnosis, prognosis, and treatment selection. In practice, interpretation focuses on features associated with the most prominent waves of each heartbeat, commonly labeled P, Q, R, S, and T \cite{de2008basic}.

The ECG data $X(\boldsymbol{t})$ is the voltage measured from a heartbeat with a single electrode, with the time scale modified to be in $[0,2\pi)$. The FMM$_{ecg}$ model for $X(\boldsymbol{t})$ is defined as follows:

$$ X(t) = M+\sum_{k\in\{P,Q,R,S,T\}} A_kW(t;\alpha_k,\beta_k,\omega_k)+\epsilon(t)$$   
$$ \alpha_{P}  \le \alpha_{Q} \le \alpha_{R} \le \alpha_{S}  \le \alpha_{T}$$

The FMM$_{ecg}$ model is physiologically interpretable, as the components align with the five main waves in the ECG, each representing the electrical activity generated in different regions of the heart. Specifically, the constraints on the parameters $\alpha$'s correspond to the propagation of the electrical signal across the heart, from the sinus node to the ventricles \cite{franzone2014mathematical}.
The estimation algorithm incorporates an identification step where components are labeled as PQRST. Moreover, for pathological hearts or very noisy signals other restrictions among parameters are included to improve the identification. The FMM$_{ecg}$ model was initially developed for unidimensional signals in \cite{rueda2021hidden} and extended to the multi-channel case, the 3DFMM$_{ecg}$ model, where the three-dimensional structure of the signal enhances the algorithm's robustness by providing additional physical interpretations of the parameters \cite{rueda2022unique}.
On the other hand, we also propose an  AFD approximation with 5 components to analyze $X(t)$.

The applications of AFD for ECG analysis primarily focus on signal compression and denoising. In such cases, the number of components depends directly on the desired Signal-Noise Ratio (SNR) level. The number of components can be as high as 35 when analyzing ECG signals \cite{compressionECGma2014novel, compressionECGbanerjee2022, ECGdenoisingWang}.\\

To illustrate the differences between FMM$_{ecg}$ and AFD decompositions, we have selected two heartbeats from the PTBXL, a benchmark database in ECG analysis (see \cite{PTBXL2020}), with different patterns. The first is a typical pattern from a healthy heart, while the second shows an anomalous T-wave morphology from a non-healthy heart. Additionally, the second has a lower SNR compared to the first one.

\begin{figure*}[ht!] 
    \centering
    \includegraphics[width=\textwidth]{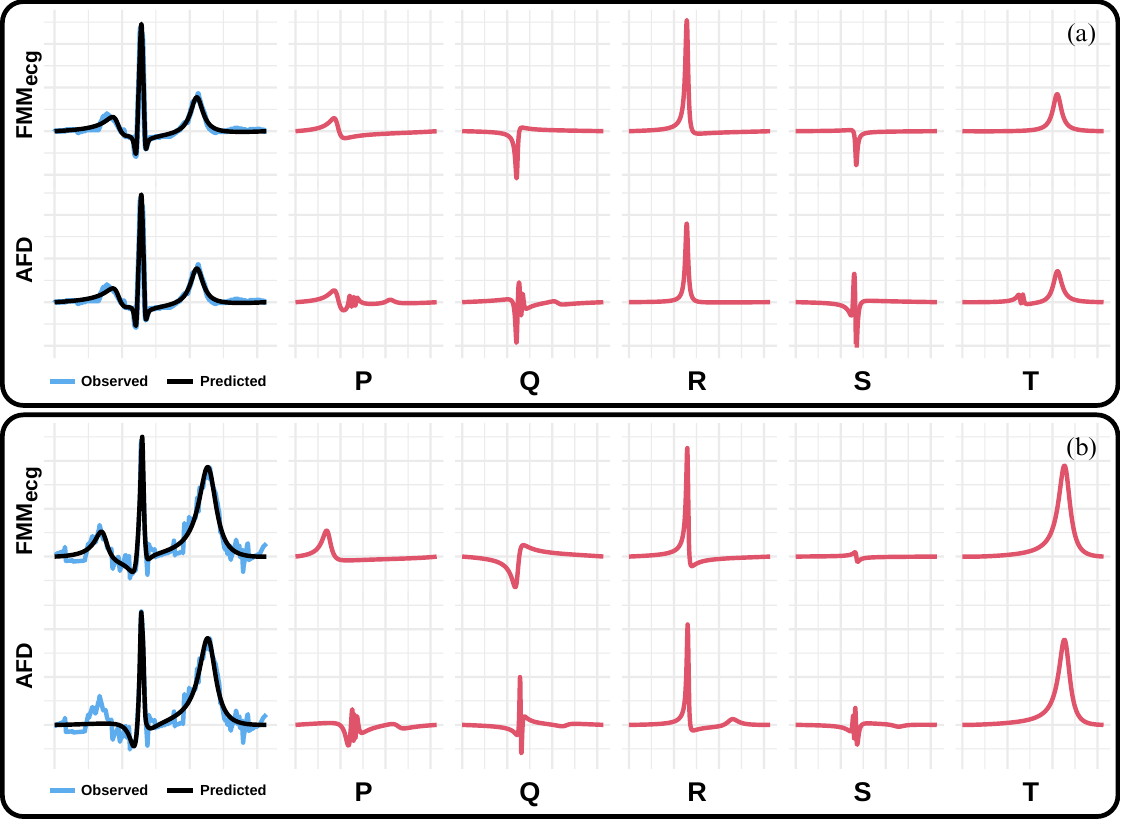}
    \caption{ECG signal analysis for heartbeats corresponding to patients 13022 (panel (a)) and 1119 (panel (b)) from the PTBXL database. In each panel we display, in column 1: in blue, the observed signal from lead II; in black, the  FMM$_{ecg}$ (top row) and AFD (bottom row)  predictions. In columns 2 to 6 the FMM$_{ecg}$ (top row) and AFD (bottom row) are real parts of the individual components. The components are sorted by the FMM $\alpha$'s parameters that correspond to the PQRST sequence.
    } \label{ecgExample}
\end{figure*}

The FMM estimates are derived using the model FMM$_{ecg}$ software. The AFD parameter estimates are obtained by leveraging FMM parameter estimates, which are obtained using the R package FMM, as well as Theorem~\ref{Theorema}. These AFD estimates align with those computed using the accurate cyclic AFD method discussed in Section~\ref{algorithms}. The components extracted using AFD are not initially labeled.

In Figure~\ref{ecgExample},  panel (a) corresponds to the first and panel (b) to the second. In the first column of both panels, the blue line represents the observed ECG signal from lead II and the black line shows the prediction using the real parts of the  FMM$_{ecg}$ (top row) and AFD (bottom row) five-component decompositions. Furthermore, for each example, the patterns of the real parts of the individual components are displayed in columns 2 to 6, sorted by the FMM $\alpha$'s parameters, which correspond to the PQRST sequence.

In the example in panel (a), the ECG morphology is correctly delineated with both FMM$_{ecg}$ and AFD. However, while the FMM$_{ecg}$ components each describe one of the PQRST waves, this is not the case for AFD. For instance, all five AFD components participate in describing the most prominent peak, the R wave.
In the example in panel (b), the FMM$_{ecg}$ still accurately describes the five prominent peaks of the signal, whereas the AFD fails to delineate the first peak, corresponding to the P wave. The individual AFD components do not align with the known ECG waves in a direct manner and differ from those in the first case, as the algorithm does not incorporate an identification step. Without the identification step, the FMM would provide similar results, as stated in the experiments in Section \ref{algorithms}.

The parametrization FMM$_{ecg}$ aids the interpretation process by identifying distinct features of individual waves using wave-specific parameters, thereby emulating the way physicians analyze ECGs. Therefore, it is recommended when the study objective involves developing diagnostic rules, classification or pattern recognition, among other applications. Furthermore, it is also beneficial for suppressing noisy components characterized by extreme parameter values. Specifically, components with $\alpha$ values too close to the beginning or end of the heartbeat, delineate peaks that may correspond to the anterior or posterior heartbeat, and those with too low $\omega$ values describe very sharp peaks, such as those generated by pacemaker or other artifacts.

\section{Discussion and conclusions} \label{Discussion}

AFD and FMM approaches have distinct origins but share the same underlying complex-domain framework. AFD extends Fourier analysis by enabling flexible positive-frequency decompositions and has been recognized as an effective greedy strategy, with important advantages for computation and for the design of efficient optimization algorithms. In contrast, FMM is formulated as a statistical model aimed at extracting interpretable features from real signals. A key strength of FMM is its ability to describe prominent oscillations with single components, which facilitates component-wise interpretation in real-world applications; however, this non-orthogonal parametrization may increase the difficulty of numerical optimization.
\\

A main implication of the equivalence established in this work is that AFD and FMM can be related in a way that enables methodological transfer between both perspectives. In particular, FMM can benefit from AFD's computational efficiency (including fast approximation machinery), while AFD can benefit from FMM's interpretability and statistical tools. This is especially relevant in ECG signal processing, where identifying physiologically meaningful components and separating noisy elements are important for downstream analyses such as waveform characterization and feature extraction.
\\

Overall, the results provide a principled basis to relate two widely used yet largely separate approaches to oscillatory signal decomposition. By establishing finite-order equivalence and the corresponding equivalence of estimation problems under independent Gaussian errors, the paper clarifies when AFD and FMM can be expected to deliver comparable approximations with a fixed number of components.

From a biomedical signal-processing perspective, the key practical implication is that one may leverage fast AFD approximation machinery (including FFT-based implementations) while retaining FMM-style parametrization when component-level interpretability or identification is required, as illustrated in the ECG use case. Importantly, even when fitted signals coincide, interpretability at the component level can differ, since AFD components depend on the ordering of the selected parameters and may mix oscillatory patterns in noisy scenarios.

In future work, we will develop hybrid procedures that explicitly exploit this connection, using AFD-based searches to accelerate and stabilize estimation of nonlinear parameters and subsequently relying on FMM parametrization for interpretable component extraction. We will also introduce inference tools and explore extensions beyond the independent Gaussian error setting, including nonlinear or correlated random terms and practically motivated parameter constraints.

\section*{Acknowledgement}

This work was supported by the Spanish Ministry of Science, Innovation and Universities under Grant PID2023-147839OB-I00 (to C.R.); and by the Spanish Ministry of Science and Innovation, and the European Regional Development Fund (ERDF) program of the European Union under Grant PID2022-142292NB-I00 (NATASHA Project, to R.C.-S.).

\bibliographystyle{elsarticle-num}
\bibliography{bibliography}

\end{document}


\maketitle
\vspace{-1cm} 
\begin{center}
\textsuperscript{1} Department of Statistics and Operative Research, University of Valladolid,\\ Valladolid, Spain \\
\textsuperscript{2} Department of Informatics, University of València,\\ Valencia, Spain
\end{center}

\noindent
\setlength{\parindent}{0pt} 
\setlength{\parskip}{1ex plus 0.5ex minus 0.5ex} 

This supplementary material provides proofs, algorithm descriptions, and additional details for replicating the methods and results from the main text. It includes the proofs of Lemmas 1, 2, and Theorem 3, as well as algorithmic descriptions of the FMM (Frequency-Modulated Möbius) and AFD (Adaptive Fourier Decomposition) methods.

The material is organized as follows: Section 1 presents the proofs of the lemmas and the theorem, while Section 2 outlines the FMM and AFD algorithms, focusing on their differences.

\setlength{\parindent}{0pt}

\section{Proofs of Lemmas 1, 2 and Theorem 3.}

\subsection{Proof of Lemma 1}

Let \( m_{a_1}(z) \) and \( m_{a_2}(z) \) be two Möbius transformations with \( a_1 \neq a_2 \). We aim to express the product \( m_{a_1}(z) m_{a_2}(z) \) as a linear combination of the form:

$$ m_{a_1}(z) m_{a_2}(z) = \lambda_0 + \lambda_1 m_{a_1}(z) + \lambda_2 m_{a_2}(z), $$
which we derive by first obtaining a common denominator on the right-hand side and focusing on the numerators. This leads to the following equation:

$$(z-a_1)(z-a_2) = \lambda_0(1-\bar{a}_1z)(1-\bar{a}_2z) + \lambda_1(z-a_1)(1-\bar{a}_2z) + $$
$$+ \lambda_2(z-a_2)(1-\bar{a}_1z).$$

Expanding this expression results in the following system of equations:

\begin{equation*} \label{Eq:Sys}
\left\{
\setlength{\arraycolsep}{0pt}%
  \renewcommand{\arraystretch}{1.2}%
  \begin{array}{ *{5}{ l >{{}} l <{{}} } r }
  
  & \lambda_0 - \lambda_1a_1 - \lambda_2a_2 = a_1a_2\\
  & \lambda_0(\bar a_1+\bar a_2) - \lambda_1(1+ a_1\bar a_2) -\lambda_2(1 + a_2\bar a_1) = a_1+a_2\\ 
  & \lambda_0\bar{a}_1\bar{a}_2-\lambda_1\bar{a}_2 - \lambda_2\bar{a}_1 = 1,
\end{array}
\right.
\end{equation*}

The solutions to this system are:

$$ \lambda_0 = \frac{a_1 - a_2}{\bar{a}_1 - \bar{a}_2}, \quad \lambda_1 = \frac{1 - \bar{a}_1 a_2}{\bar{a}_1 - \bar{a}_2}, \quad \lambda_2 = \frac{1 - \bar{a}_2 a_1}{\bar{a}_2 - \bar{a}_1}.$$

Note that \( \lambda_0 \) is independent of the order of \( a_1 \) and \( a_2 \), and that \( \lambda_1(a_1, a_2) = \lambda_2(a_2, a_1) \).

\subsection{Proof of Lemma 2} 

We proceed by induction to prove the result for the product of \( k \) Möbius transformations. The base case of the induction corresponds to Lemma 1, which shows that the product of two Möbius transformations can be expressed as a linear combination of Möbius transforms with specific coefficients. In this proof, we aim to extend this result to the product of \( k \) Möbius transformations, giving analytic expressions for the coefficients.

At the \( k \)-th step, we assume that the product of the first \( k-1 \) Möbius transformations can be written as a sum of Möbius transforms with coefficients \( \phi_0^{k-1}, \phi_1^{k-1}, \dots, \phi_{k-1}^{k-1} \) and a free term \( \phi_0^{k-1} \), i.e.,

$$\prod_{i=1}^km_i(z) = \left(\prod_{i=1}^{k-1}m_i(z)\right)m_k(z).$$ 
Now, applying the induction hypothesis: the product for $m_1(z)...m_{k-1}(z)$ can be expressed as a sum with coefficients $\phi_1^{k-1},...,\phi_{k-1}^{k-1}$ and a free term $\phi_0^{k-1}$,

$$\left(\phi_0^{k-1} + \sum_{i=1}^{k-1}\phi_i^{k-1}m_i(z)\right)m_k(z) =\phi_0^{k-1}m_k(z) + \sum_{i=1}^{k-1}\phi_i^{k-1}m_i(z)m_k(z). $$

By applying Lemma 1 to the product \( m_i(z) m_k(z) \), we can express each product \( m_i(z) m_k(z) \) as a linear combination of Möbius transforms. 

$$\prod_{i=1}^km_i(z) =\phi_0^{k-1}m_k(z) + \sum_{i=1}^{k-1}\phi_i^{k-1}\lambda_0^{(i,k)} + \sum_{i=1}^{k-1}\phi_i^{k-1}\lambda_1^{(i,k)}m_i(z)+ $$
$$  + \sum_{i=1}^{k-1}\phi_i^{k-1}\lambda_2^{(i,k)}m_k(z) $$

where the expressions for $\phi_0^k,...,\phi_k^k$ are obtained after grouping the coefficients for each Möbius transform, and $\lambda_*^{(i,j)}$ represents the coefficients of the linear combination in Lemma 1 given $a_i$ and $a_j$, where $i\neq j$.

\subsection{Proof of Theorem 3}

Let \( a_1 \neq \cdots \neq a_K \in \mathbb{D} \), and let \( \{B_{k}\}_{k=0}^K \) and \( \{m_{a_k}\}_{k=0}^K \) be the associated Takenaka-Malmquist and Möbius bases, respectively. 

To show that

$$ F(z) = c_0 + \sum_{k=1}^K c_k B_k(t) = \phi_0 + \sum_{k=1}^K \phi_k m_{a_k}(t) \in L^2(\partial\mathbb{D}), $$

we first proceed to develop each AFD component:

$$
c_k B_k = c_k z \frac{\sqrt{1 - |a_k|^2}}{1 - \bar{a}_k z} \prod_{j=1}^{k-1} m_j(z)
$$

as follows:

\begin{equation} \label{eq:CkBk}
    \frac{c_k}{\sqrt{1 - |a_k|^2}} \left( a_k \prod_{j=1}^{k-1} m_j(z) + \prod_{j=1}^{k} m_j(z) \right).
\end{equation}

Next, we apply Lemma 2 to equation \eqref{eq:CkBk} to obtain:

$$\frac{c_k}{\sqrt{1-|a_k|^2}}\left(a_k\phi_0^{k-1}+\phi_0^k\right)+\frac{c_k}{\sqrt{1-|a_k|^2}}\left(\sum_{i=1}^{k-1}\frac{1-|a|^2}{\bar{a}_i-\bar{a}_k}m_i(z)\right) + \frac{c_k}{\sqrt{1-|a_k|^2}}\phi_k^km_k(z).$$

We group the coefficients for each \( m_k(z) \) for $k=1,...,K$. This leads to expressions for \( \phi_0, \phi_1, \dots, \phi_K \), which can be represented by constructing the following matrix from the expressions:

\begin{equation*} 
\mathcal{M}_{\boldsymbol{a}} = \begin{pmatrix}
1 & \frac{a_1}{\sqrt{1 - |a_1|^2}} & \dots & \frac{a_k \xi_0^{k-1} + \xi_0^k}{\sqrt{1 - |a_k|^2}} & \dots &  \frac{a_K \xi_0^{k-1} + \xi_0^k}{\sqrt{1 - |a_K|^2}} \\
0 & \frac{1}{\sqrt{1 - |a_1|^2}} & \dots & \frac{\xi_1^k \sqrt{1 - |a_k|^2}}{\overline{a_1} - \overline{a_k}} & \dots &  \frac{\xi_1^K \sqrt{1 - |a_K|^2}}{\overline{a_1} - \overline{a_K}} \\
\vdots & \vdots & \ddots & \vdots & \ddots & \vdots \\
0 & 0 & \dots & \frac{\xi_k^k}{\sqrt{1 - |a_k|^2}} & \dots & \frac{\xi_k^K \sqrt{1 - |a_K|^2}}{\overline{a_k} - \overline{a_K}} \\
\vdots & \vdots & \ddots & \vdots & \ddots & \vdots \\
0 & 0 & \dots & 0 & \dots & \frac{\xi_K^K}{\sqrt{1 - |a_K|^2}}
\end{pmatrix},
\end{equation*}

where each \( \phi_k \) is the dot product of the corresponding row of the matrix and the vector \( (c_0, c_1, \dots, c_K)^T \), and $\boldsymbol{a}$ denotes $(a_1,...,a_K)$. The matrix \( \mathcal{M}_{\boldsymbol{a}} \) is triangular and invertible when \( a_1 \neq \dots \neq a_K \), defining the change-of-basis with the coordinates associated as:

\[
(\phi_0, \phi_1, \dots, \phi_K)' = \mathcal{M}_{\boldsymbol{a}} \, (c_0, c_1, \dots, c_K)'.
\]

Thus, this change-of-basis matrix establishes a one-to-one correspondence between the AFD and FMM coefficients, proving that both decompositions represent the same subspace of \( L^2(\partial \mathbb{D}) \) signals.

The matrix $\mathcal{M}$ directly relate the AFD and FMM coefficients. Another way to see the correspondence between the theoretical decomposition is to note that the linear span of a Möbius basis \(\{m_{a_k}\}_{k=0}^K \) with mutually different $a$' s is equivalent to the linear span of a Szego kernel basis \(\{e_{a_k}\}_{k=0}^K \) with the same $a$'s. Since the T-M system \( \{B_{k}\}_{k=0}^K \) is the Gram-Schmidt orthogonalization of \(\{e_{a_k}\}_{k=0}^K \) as stated in \cite{QianRemarks2013}, the correspondence is valid.

\section{FMM and AFD algorithms}

In this section, we provide a detailed overview and comparison of the key algorithms used in the FMM (Frequency-Modulated Möbius) and AFD (Adaptive Fourier Decomposition) methodologies. We begin by describing the FMM backfitting algorithm, which is central to the FMM framework, followed by the core AFD algorithm and its cyclic extension. 

\subsection{FMM Backfitting algorithm} \label{subsec:D}

In this section, we describe the FMM backfitting algorithm, which is the foundational algorithm in the FMM framework \cite{FMMpackage}. After fitting the initial \( K \) components, the algorithm proceeds by iterating, as outlined in Algorithm~\ref{algorithm1}, until a specified convergence criterion is met. Once the best values for \( (\alpha_k, \omega_k) \in \mathcal{D} \) are identified, the solution is further refined using an optimization routine. Common optimization methods such as the Nelder-Mead method \cite{nelder-mead} or a gradient-based approach can be employed. These routines take advantage of the smoothness and convexity properties of the likelihood function to locate critical points outside of the initial grid, reducing reliance on the grid density \( \mathcal{D} \) and enabling the identification of critical values without the need for a finely tuned grid.

\begin{algorithm} 
  \caption{FMM backfitting algorithm} \label{algorithm1}
  \begin{algorithmic}[1]
    \Require{$\boldsymbol{t},X(\boldsymbol{t}), MaxIter,K,\mathcal{D}$} 
    
    \State{$\mu_1^{(0)}(\boldsymbol{t})=...=\mu_k^{(0)}(\boldsymbol{t}) \gets 0$} 
    
    \For{$i = 1,\dots,MaxIter$}
        \For{$k = 1,\dots,K$}
        \State{$r^{(i)}_k(\boldsymbol{t}) \gets X(\boldsymbol{t})-\sum_{j<k} \mu_k^{(i)}(\boldsymbol{t})-\sum_{j>k} \mu_k^{(i-1)}(\boldsymbol{t})$ }
        \State{$(\tilde{\alpha},\tilde{\omega})^{(i)}_{k} \gets \argmin_{\alpha,\omega\in\mathcal{D}}\| r^{(i)}_k(\boldsymbol{t})-\mu_k(\boldsymbol{t}) \|^2$}  
        \State{$(\hat{\alpha},\hat{\omega})^{(i)}_{k} \gets Optim( (\tilde{\alpha},\tilde{\omega})^{(i)}_{k}, \| r^{(i)}_k(\boldsymbol{t})-\mu_k(\boldsymbol{t}) \|^2$}) 
        \EndFor
    \EndFor
  \end{algorithmic}
  \Return{$(\hat{\alpha},\hat{\omega})^{(MaxIter)}_{1},\dots,(\hat{\alpha},\hat{\omega})^{(MaxIter)}_{K}$}
\end{algorithm}

This algorithm is designed to find a critical point in the global optimization problem. Additionally, the flexibility of the FMM framework allows for the incorporation of parameter constraints, which are often informed by physiological considerations. This adaptability makes the FMM particularly suitable for adjustments tailored to specific problems.

The FMM software has been developed in R, with a package available for the univariate case \cite{FMMpackage}, as well as GitHub repositories\footnote{\url{https://github.com/FMMGroupVa}} for the multivariate case. Several FMM-based algorithms have been implemented for bioelectric signal analysis, including classification, identification, and inference tools.

\subsection{Core AFD and Cyclic AFD algorithms} \label{subsec:E}

The most basic AFD algorithm, which follows the iterative scheme introduced in the previous section, is known as the AFD Core algorithm \cite{QianCore}. It is described in Algorithm~\ref{algorithm2}. When combined with the FFT enhancements \cite{QianFFT}, this algorithm results in a highly efficient estimation approach for signal decomposition.

\begin{algorithm} 
  \caption{AFD core algorithm} \label{algorithm2}
  \begin{algorithmic}[1]
    \Require{$\boldsymbol{t},Z(\boldsymbol{t}), K,\mathcal{D}$} 
    \State{$z \gets e^{i\boldsymbol{t}}$}
    \State{$a_0 \gets 0$}
    \State{$\,\hat{c}_0 \gets mean(Z(\boldsymbol{t}))$}
    \State{$H_0 \gets (Z-c_0)/z$}
    \For{$k = 1,\dots,K$}
        \State{$\hat{a}_k \gets \argmax_{a\in\mathcal{D}} |\langle H_{k-1}, e_{a_k}\rangle|^2$} 
        \State{$H_k \gets (H_{k-1}-\langle H_{k-1}, e_{\hat{a}_k}\rangle e_{\hat{a}_k})\frac{1-\Bar{\hat{a}}_k z}{z-\hat{a}_k}$}
    \EndFor
  \end{algorithmic}
  \Return{$\hat{a}_1,\dots,\hat{a}_K$}
\end{algorithm}

Similar to the backfitting algorithm, the cyclic AFD \cite{QianCyclic} has been proposed as an enhancement to improve the optimization process. This method ensures convergence to a critical point of the objective function, providing a more efficient solution compared to the basic AFD algorithm.

The AFD software has been primarily developed in MATLAB and Python (for more details, visit the toolbox website\footnote{\url{https://toolbox-for-adaptive-fourier-decomposition.readthedocs.io}}). 

In contrast to FMM, which leverages unique oscillatory components that facilitate the straightforward incorporation of physiological constraints, AFD's inherent orthogonality structure imposes a more rigid framework, limiting the flexibility in introducing problem-specific constraints.

\bibliographystyle{plain}
\bibliography{bibliography}